\documentclass[prl,reprint]{revtex4-1}
\usepackage{amsmath}
\usepackage{multirow}
\usepackage{bm}
\usepackage{braket}
\usepackage{graphicx}
\usepackage{dcolumn}

\begin{document}

\title{Efficient Tree Tensor Network States (TTNS) for Quantum Chemistry: Generalizations of the Density Matrix Renormalization Group Algorithm}

\author{Naoki Nakatani}
\author{Garnet Kin-Lic Chan}
\affiliation{Department of Chemistry, Princeton University, Frick Chemistry Laboratory, Princeton, NJ 08544, USA}

\date{\today}

\begin{abstract}
We investigate tree tensor network states for quantum chemistry.
Tree tensor network states represent one of the simplest generalizations
of matrix product states and the density matrix renormalization group.
While matrix product states encode a one-dimensional
entanglement structure, tree tensor network states encode 
a tree entanglement structure, allowing for a more flexible description of general molecules.
We describe an optimal tree tensor network state
algorithm for quantum chemistry. We introduce the concept of half-renormalization
which greatly improves the efficiency of the calculations.
Using our efficient formulation we demonstrate
the strengths and weaknesses of tree tensor network states versus matrix product states.
We carry out benchmark calculations both on tree systems (hydrogen trees and $\pi$-conjugated
dendrimers) as well as non-tree molecules (hydrogen chains, nitrogen dimer, and chromium dimer).
In general, tree tensor network states require much fewer renormalized states
to achieve the same accuracy as matrix product states. 
In non-tree molecules, whether this translates into a computational savings is system dependent,
due to the higher prefactor and computational scaling associated with tree algorithms.
In tree like molecules, tree network states are easily superior to matrix product states. As
an illustration, our largest dendrimer calculation with tree tensor network states 
correlates 110 electrons in 110 active orbitals.
\end{abstract}

\maketitle

\section{Introduction}

Currently, there is much effort devoted to finding efficient numerical techniques for
 strongly correlated electrons.
Amongst several approaches, the Density Matrix Renormalization Group (DMRG)\cite{white1992,white1993} has provided
many new insights in challenging systems. In recent years, efficient
DMRG implementations have also appeared for quantum chemistry.\cite{white1999,chan2002,sharma2011,legeza2003a,reiher2007,reiher2010,kura2009}
Originally, the DMRG was formulated as an algorithm in the language of Wilson's numerical renormalization group.
However, more recently, attention has expanded to the underlying class of wavefunctions optimized by the DMRG,
which are the Matrix Product States (MPS).


MPS provide a compact description of entanglement in 1D systems,  thus
DMRG calculations for chain-like molecules are very efficient.\cite{chan2006,chan2008}
However, when used in 2D and 3D systems,
 much larger bond-dimensions, usually denoted by $M$ and referred to as the number of renormalized states in 
DMRG calculations, must be used to reach a target accuracy. The need for larger $M$ reflects
 the sub-optimal representation of 2D and 3D entanglement by the MPS wavefunction structure.


The generalization of MPS to tensor network states (TNS)  provides a natural way to compactly describe
2D and 3D entanglement. There are several families of TNS that differ in
the way the entanglement is encoded, and a brief overview of the classes of TNS is given in Refs.~\onlinecite{vidal2006}--\onlinecite{chan2012}.
While the formal properties of general TNS are appealing, efficient computation with these states
lags far behind computation with MPS. In this work, we explore efficient computation for quantum chemistry
with Tree Tensor Network States (TTNS),\cite{vidal2006,luca2009,murg2010,xiang2012,hitesh2013} one of the simplest families of TNS.



\begin{figure}
\includegraphics[scale=0.8]{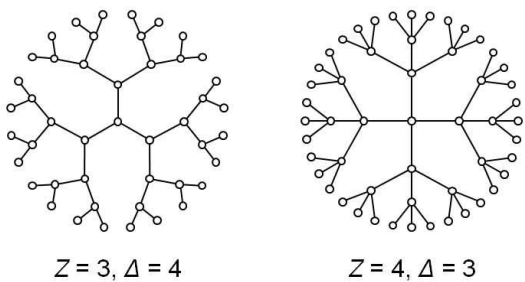}
\caption{\label{fig:bethe_tree} Examples of trees.
         The left panel shows a tree with degree $Z=3$ and depth $\Delta=4$, and
         the right panel shows a tree with degree $Z=4$ and depth $\Delta=3$.}
\end{figure}

TTNS encode a tree entanglement structure, as illustrated in Figure~\ref{fig:bethe_tree}.
A tree of maximal degree $Z$ has at most $Z$ neighbours at any site. MPS are a special
case of TTNS with $Z=2$. The absence of cycles in a TTNS simplifies many algorithms. 
In particular, the variational minimization of TTNS wavefunctions can be carried out by generalizing
the DMRG algorithm used for MPS, from a two-block formulation, to a $Z$-block formulation. Multi-block DMRG
algorithms have been studied for some time,\cite{otsuka1996,friedman1997,bursill1999,lepetit2000,sierra2002,soos2012} although to the best of our knowledge,
their interpretation in terms of the underlying class of TTNS variational states first
appeared in Ref.~\onlinecite{murg2010}. Thus, Ref.~\onlinecite{murg2010} can be considered as a starting point for
the current work.

In Ref.~\onlinecite{murg2010} the authors considered a prototype application of TTNS
to quantum chemistry with a minimal basis beryllium atom calculation.
Although this provided  evidence that for a given $M$, 
TTNS of degree $Z>2$  capture more entanglement than the corresponding MPS with the same $M$,
for practical computation many questions were left unanswered. For example, 
although TTNS are more flexible than MPS for describing entanglement, there is an implicit trade-off of 
higher computational scaling with $M$.
Furthermore, the calculation in Ref.~\onlinecite{murg2010} used an unrealistically small $M=3$, whereas typical
DMRG calculations use $M=1000-10000$. Thus, the relevance of TTNS
for quantum chemistry calculations remains to be established.

In the current work, we address this question by describing and implementing an efficient 
TTNS algorithm that can be used for realistic calculations. 
Our algorithm is formulated to have an optimal computational scaling for quantum chemistry
Hamiltonians. We achieve a much lower cost than
earlier formulations through a ``half''-renormalization transformation 
which exactly maps the multi-block DMRG  to a conventional, efficient, two-block DMRG.
We also address the issue of orbital ordering on trees which is necessary to use TTNS in chemistry.
Using our efficient implementation, we  assess the performance of TTNS relative to MPS used
in quantum chemistry DMRG calculations.
We compare TTNS and MPS on several benchmark molecular systems, including ideal hydrogen trees and chains,
and benchmark molecules such as the nitrogen and chromium dimers.
Finally, to demonstrate the power of TTNS, we carry out complete active space calculations 
in $\pi$-conjugated dendrimer molecules, correlating up to 110 electrons in 110 orbitals.



\section{Overview of the DMRG algorithm based on MPS}

We first present a brief overview of MPS wavefunctions and the
DMRG algorithm that optimizes their energy.
This will allow us to establish basic notation which will be used
to discuss TTNS in the next section. However, 
as the presentation is not entirely self-contained, for further details 
we refer to additional articles and reviews.\cite{white1992,chan2002,cirac2008,schollwock2005,schollwock2011}

We use the following notation for MPS: $n_{i}$ is the many-body basis at site \textit{i}, 
\textit{k} is number of sites, and \textit{M} is the number of renormalized states.
The (one-site) MPS wavefunction is obtained by expressing the 
coefficient of the determinant  $\ket{n_1...n_k}$ as a product of matrices
for each occupancy $n_1...n_k$,
\begin{equation} \label{eq:dmrg_wfu_mps}
  \ket{\Psi}
 =\sum_{n_{1}...n_{k}} \mathbf{L}^{n_{1}}...\mathbf{L}^{n_{i-1}} \bm{\psi}^{n_{i}}
                       \mathbf{R}^{n_{i+1}}...\mathbf{R}^{n_{k}} \ket{n_{1}...n_{k}}
\end{equation}
For an MPS with $M$ renormalized states, the matrices are of maximum dimension $M \times M$,
except for the first and last, which are of maximum dimension $1 \times M$ and $M \times 1$ respectively.
Note that the MPS is invariant to a number of transformations of the matrices.\cite{schollwock2011,sharma2011}
We remove this invariance by choosing an MPS {\it canonical form}. In the canonical form at site $i$, the rotation matrices to
the left  of site $i$ are constrained to satisfy 
orthonormality conditions $\sum_{n_{i}} \mathbf{L}^{n_{i}\dag} \mathbf{L}^{n_{i}} = \mathbf{1}$,
while those to the right of site $i$ are constrained to satisfy
 $\sum_{n_{i}} \mathbf{R}^{n_{i}} \mathbf{R}^{n_{i}\dag} = \mathbf{1}$. 



The left and right rotation matrices allow us to define left and right renormalized many-body states,
 $\ket{l_{i-1}}$ and $\ket{r_{i}}$, respectively. These renormalized representations
are used to construct the computational intermediates (renormalized operators) in the 
DMRG algorithm.
Carrying out the matrix multiplications from $\mathbf{L}^{n_{1}}$ through $\mathbf{L}^{n_{i-1}}$,
and from $\mathbf{R}^{n_{k}}$ through $\mathbf{R}^{n_{i+1}}$, we obtain
\begin{align}
  \ket{l_{i-1}}
&=\sum_{n_{1}...n_{i-1}} \mathbf{L}^{n_{1}}...\mathbf{L}^{n_{i-1}} \ket{n_{1}...n_{i-1}} \\
  \ket{r_{i}}
&=\sum_{n_{i+1}...n_{k}} \mathbf{R}^{n_{i+1}}...\mathbf{R}^{n_{k}} \ket{n_{i+1}...n_{k}}.
\end{align}
The orthonormality conditions on the rotation matrices $\mathbf{L}$ and $\mathbf{R}$ imply that
the renormalized bases are orthonormal
\begin{align}
\braket{l_{i-1}|l_{i-1}'} &= \delta_{ll'} \\
\braket{r_{i}|r_{i}'} &= \delta_{rr'} 
\end{align}

The DMRG (canonical) form of the MPS wavefunction is obtained by rewriting
the MPS wavefunction (\ref{eq:dmrg_wfu_mps}) in terms of the renormalized many-body bases  $\ket{l_{i-1}}$ and $\ket{r_{i}}$,
\begin{equation} \label{eq:dmrg_wfu_i}
  \ket{\Psi}
 =\sum_{l_{i-1}n_{i}r_{i}} \psi_{l_{i-1}r_{i}}^{n_{i}} \ket{l_{i-1}n_{i}r_{i}}.
\end{equation}
In this interpretation, $\bm{\psi}^{n_{i}}$ is viewed as a wavefunction coefficient vector 
in a Hilbert space spanned by the renormalized product states, $\ket{l_{i-1}n_{i}r_{i}} = \ket{l_{i-1}} \ket{n_{i}} \ket{r_{i}}$. 


\begin{figure*}
\includegraphics[scale=0.5]{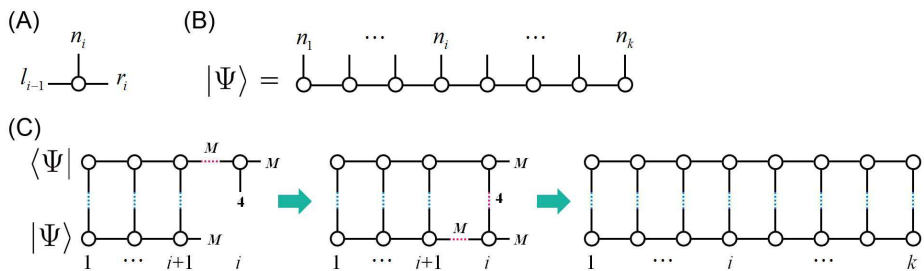}
\caption{Graphical representations of MPS.
(A) one-site coefficient vector $\bm{\psi}^{n_{i}}$ is a rank-3 tensor,
(B) the MPS wavefunction is given by contracting all horizontal bonds,
(C) the overlap of two MPS wavefunctions can be efficiently computed recursively.}
\label{fig:mps_graph_rep}
\end{figure*}

Computations using MPS involve  tensor-tensor contractions. To
express such operations, it is helpful to use a graphical representation, shown in Figure~\ref{fig:mps_graph_rep}.
Each vertex is  a tensor and the number of edges connected to the vertex determines the tensor rank. 
In the case of an MPS, each $\mathbf{L}^{n_i}$ or $\mathbf{R}^{n_i}$ is a rank-3 tensor, represented by a vertex with three edges.
Here, we always choose the vertical index to represent $n_{i}$.
The MPS wavefunction is obtained by contracting  the horizontal edges of all the tensors (Fig.~\ref{fig:mps_graph_rep} (B)), leading to
 Eq.~(\ref{eq:dmrg_wfu_mps}). The computation of the overlap of two MPS is shown in  Fig.~\ref{fig:mps_graph_rep} (C).


We optimize the MPS energy, by minimizing the Lagrangian $\braket{\Psi|\hat{H}|\Psi} - (\lambda \braket{\Psi|\Psi} - 1)$
 with respect to the tensors in the MPS. 
In the (one-site) DMRG sweep algorithm, this minimization is carried out with respect
to a single tensor at a time. In  step $i$ of the DMRG sweep, the MPS is expressed in the DMRG
form (\ref{eq:dmrg_wfu_i}), and the coefficient vector $\bm{\psi}$ is optimized, holding the rotation matrices to the
left and right of the site which define the bases $\ket{l_{i-1}}$, $\ket{r_i}$, fixed. Because the energy is a quadratic form in 
$\bm{\psi}$,  minimization  leads to a standard 
eigenvalue problem,
\begin{equation}
\sum_{l'_{i-1}n'_{i}r'_{i}}
   \braket{l_{i-1}n_{i}r_{i}|\hat{H}|l'_{i-1}n'_{i}r'_{i}} \psi_{l_{i-1}r_{i}}^{n_{i}} -\lambda \psi_{l_{i-1}r_{i}}^{n_{i}} \\
=0
\label{eq:eig}
\end{equation}
In subsequent steps of the sweep, the MPS is transformed to the DMRG form at successive sites, and the 
coefficient vectors at these sites are  optimized. 


The most expensive operation in the sweep is computing $\braket{\Psi|\hat{H}|\Psi}$ and
performing $\hat{H}\psi$ to solve the eigenvalue problem (\ref{eq:eig}).
This is because the second-quantized Hamiltonian contains a large number of terms,
\begin{equation} \label{eq:ham2nd}
  \hat{H}
 =\sum_{ij} h_{ij} \hat{a}_{i}^{\dag} \hat{a}_{j}
 +\frac{1}{2}\sum_{ijkl} v_{ijlk} \hat{a}_{i}^{\dag} \hat{a}_{j}^{\dag} \hat{a}_{k} \hat{a}_{l}.
\end{equation}
and for each term we need its matrix representation in the basis $\{ \ket{l_{i-1}n_i r_i}\}$.
There are two generic strategies to handle the large number of terms.
The first, used in existing quantum chemistry DMRG implementations, is through
 \textit{Complementary Operators}.\cite{white1999,chan2002,xiang1996} The second uses the more recent concept of \textit{Matrix Product Operators}.\cite{cirac2004,mcculloch2007,crosswhite2008,pirvu2010,dur2010}

\begin{figure}
\includegraphics[scale=0.6]{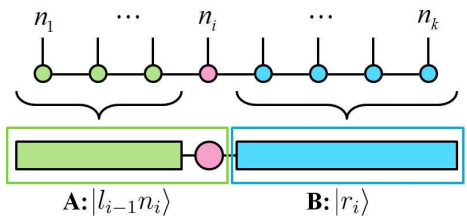}
\caption{Graphical representations of DMRG wavefunction.
Canonical form of MPS wavefunction (top) can be re-written as a block diagram in DMRG language (bottom).}
\label{fig:dmrg_graph_rep}
\end{figure}

Complementary operators are a way to  maximize the reuse of intermediates.
For example, there are $\mathcal{O}(k^4)$ terms in the summation (\ref{eq:ham2nd}), and  each
expectation value individually is of $\mathcal{O}(M^3)$ cost,  leading to a naive scaling of $\mathcal{O}(M^3 k^4)$ for
the energy. However,  much information  can be reused between terms. For example, the 
two terms $\langle a^\dag_1 a^\dag_2 a_3 a_4\rangle$ and $\langle a^\dag_1 a^\dag_2 a_6 a_7\rangle$
involve the same partial expectation value over $a^\dag_1 a^\dag_2$. 
Complementary operators reuse and combine such partial traces.
At site $i$ in the DMRG sweep, we partition the Hilbert space into two subspaces: A, containing the
left block of sites (sites $1 \ldots i-1$) and site $i$, spanned by renormalized states $\ket{l_{i-1}n_i}$, and B, 
containing the right block
of sites $i+1 \ldots k$,  and spanned by renormalized states $\ket{r_i}$. (See Fig.~\ref{fig:dmrg_graph_rep}). 
$\hat{H}$ is correspondingly partitioned as:
\begin{equation}
  \hat{H} = \hat{H}_{A} + \hat{H}_{B} + \hat{H}_{AB}
\end{equation}
$\hat{H}_{A}$ and $\hat{H}_{B}$ are act locally on A and B, respectively,
and have non-trivial expectation values with only $\ket{l_{i-1}n_i}$ and $\ket{r_i}$ separately.
$\hat{H}_{AB}$ describes the interactions between A and B, and  is given by
a sum of products acting separately on the two spaces
\begin{equation}
\begin{split}
  \hat{H}_{AB}
 =\sum_{i\in{A},j\in{B}}
 (\hat{a}_{i}^{A\dag} \hat{S}_{i}^{B}
 +\hat{a}_{j}^{B\dag} \hat{S}_{j}^{A}
 +\hat{a}_{i}^{A\dag} \hat{a}_{i}^{A} \hat{Q}_{ii}^{B}) \\
  \quad
 +\sum_{i>j\in{A}}
 (\hat{a}_{i}^{A}     \hat{a}_{j}^{A} \hat{P}_{ij}^{B\dag}
 +\hat{a}_{i}^{A\dag} \hat{a}_{j}^{A} \hat{Q}_{ij}^{B})
 + \textrm{adjoint.}
\end{split}
\end{equation}
where $\hat{P}_{ij}$, $\hat{Q}_{ij}$, and $\hat{S}_{i}$ are the \textit{Complementary Operators},
\begin{align}
  \hat{P}_{ij}^{B}
&=\sum_{kl \in B} v_{ijlk} \hat{a}_{k} \hat{a}_{l} \\
  \hat{Q}_{ij}^{B}
&=\sum_{kl \in B} (v_{ikjl} - v_{iklj}) \hat{a}_{k}^{\dag} \hat{a}_{l} \\
  \hat{S}_{i}^{B}
&=\sum_{j \in B} t_{ij} \hat{a}_{j} + \sum_{jkl \in B} v_{ijlk} \hat{a}_{j}^{\dag} \hat{a}_{k} \hat{a}_{l}.
\end{align}
Using complementary operators,
the total complexity of evaluating $\braket{\Psi|H|\Psi}$ and performing $\hat{H}\Psi$ is reduced to $\mathcal{O}(M^3k^3+M^2k^4)$ per sweep,
including the cost of constructing the components of $\hat{H}$ in the partitioned form (renormalization).\cite{chan2002}
This is the standard procedure to evaluate the energy in the DMRG algorithm.

\begin{figure}
\includegraphics[scale=0.5]{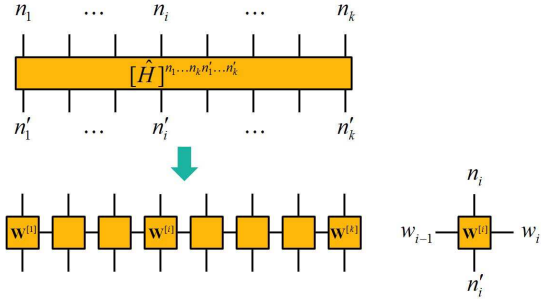}
\caption{Graphical representations of MPO.
The Hamiltonian operator is a rank-2\textit{k} tensor (top panel).
This can be divided into a contracted product of site-independent tensors similarly to an MPS,
leading to a Matrix Product Operator representation (MPO, bottom panel), in which each site tensor is a rank-4 tensor.}
\label{fig:mpo_graph_rep}
\end{figure}

An alternative approach to handle complicated Hamiltonians is through
 \textit{Matrix Product Operators} (MPO). MPO's provide a convenient
way to reason about operators in MPS algorithms and have been employed extensively in
time-dependent MPS simulations.\cite{cirac2004}
Here we provide a brief analysis of this approach for quantum chemistry.
The basic idea in an MPO is extend the matrix product like representation to operators.
Writing the Hamiltonian in the occupation number basis 
as a general rank-2\textit{k} tensor, $[\hat{H}]^{n_{1}...n_{k}n'_{1}...n'_{k}}$,
it is decomposed into a set of tensors analogously to an MPS,
\begin{equation} \label{eq:mpo}
  [\hat{H}]^{n_{1}...n_{k}n'_{1}...n'_{k}}
 =\mathbf{W}^{[1]n_{1}n'_{1}}...\mathbf{W}^{[i]n_{i}n'_{i}}...\mathbf{W}^{[k]n_{k}n'_{k}}.
\end{equation}
This is illustrated graphically in Figure~\ref{fig:mpo_graph_rep} where $\mathbf{W}^{[i]}$ is a rank-4 tensor
and the contraction of horizontal edges describes the quantum ``entanglement'' of the Hamiltonian operator.
To decompose $\hat{H}$ exactly, the dimension of the horizontal edge of $\mathbf{W}^{[i]}$
needs to be exactly the same as the  number of complementary operators, that is $\mathcal{O}(k^2)$.
Consequently, the cost of computing the expectation value $\braket{\Psi|H|\Psi}$ with an MPO
representation becomes $\mathcal{O}(M^3k^3+M^2k^5)$.
Note that this cost is {\it larger} than in the complementary operator approach. 
The difference arises because we have not considered the
sparsity of the individual $\mathbf{W}^{[i]}$ tensors that arise in the Hamiltonian decomposition.
However, incorporating element-wise sparsity into an MPO algorithm eliminates much of the conceptual and
algorithmic simplicity of the MPO approach. 
Consequently, in our view, the  complementary operator algorithm
is a more practical and efficient route for MPS computations with quantum chemistry Hamiltonians. 
The relative benefit of using complementary operators versus tensor product operators (TPO's) is even greater 
for TTNS than for MPS, 
thus we focus on the complementary operator approach when considering trees.

\section{Tree Tensor Network States (TTNS)}

Tensor Network States (TNS) are mathematical generalizations of the MPS that
can code more general entanglement networks.
The form of a TNS wavefunction is directly analogous to an MPS wavefunction,
\begin{equation} \label{eq:ttns}
  \ket{\Psi}
 =\sum_{n_{1}...n_{k}} \mathrm{ttr} \left[ \mathbf{A}^{n_{1}} \cdot ...\mathbf{A}^{n_{i}} \cdot ...\mathbf{A}^{n_{k}} \right] \ket{n_{1}...n_{k}},
\end{equation}
the only difference being that $\mathbf{A}^{n_{i}}$ is now a tensor, rather than a rotation matrix as in the MPS,
and the  multiplication operator $\cdot$ together with $\mathrm{ttr}$ denotes a general  contraction over tensor indices.
The flexibility of TNS wavefunctions arises from the fact that whereas
the matrices in an MPS can only be contracted along a 1{D}-lattice,
there are many different ways to connect general tensors together to form a network of entanglement.

\begin{figure}
\includegraphics[scale=0.5]{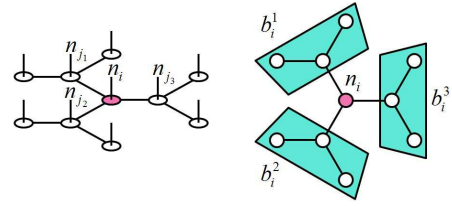}
\caption{Graphical representation of TTNS.
         Left panel shows overall structure of TTNS as described in Eq.~(\ref{eq:ttns}), and
         right panel shows the one-site wavefunction spanned by the renormalized basis, as described in Eq.~(\ref{eq:ttns_1site}).
         Note that physical indices (vertical bonds) are omitted in the right panel.}
\label{fig:ttns_graph_rep}
\end{figure}

{\it Tree tensor network states (TTNS)} are a special class of TNS where the tensors are connected as a tree 
as shown in Figure~\ref{fig:ttns_graph_rep}. A tree is a graph that has no loops,
which leads to many simplifying mathematical properties that parallel those of a MPS (indeed, an MPS is simply
a tree with $Z=2$ legs). For example, 
at a given site $i$ in the tree, we can define  renormalized bases for each of the $Z$ legs connected
to the site. This allows use to rewrite (\ref{eq:ttns}) as
\begin{equation} \label{eq:ttns_1site}
  \ket{\Psi}
 =\sum_{b_{i}^{1}...b_{i}^{Z}n_{i}} \psi_{b_{i}^{1}...b_{i}^{Z}}^{n_{i}} \ket{b_{i}^{1}...b_{i}^{Z}n_{i}}
\end{equation}
where $\ket{b_{i}^{\alpha}}$ is the renormalized basis in the $\alpha$-th branch of site $i$. This basis
is defined by recursively contracting tensors in the branch from the leaves up to site $i$. 
Thus $\ket{b_{i}^{\alpha}}$ is obtained as
\begin{equation}
  \ket{b_{i}^{\alpha}}
 =\sum_{b_{j}^{1}...b_{j}^{Z-1}n_{j}} A_{b_{j}^{1}...b_{j}^{Z-1}b_{i}^{\alpha}}^{n_{j}}
  \ket{b_{j}^{1}...b_{j}^{Z-1}n_{j}}
\end{equation}
where sites $j$ are adjacent to $i$ in the branch.
Note that for a $Z$ degree TTNS with $M$ renormalized
states, the tensor $A_{b_{j}^{1}...b_{j}^{Z-1}b_{i}^{\alpha}}^{n_{j}}$ has $\mathcal{O}(M^Z)$ elements.

Analogous to the rotation matrices in MPS, the tensors $\mathbf{A}^{n_{i}}$ in a given branch around site $i$ can be 
chosen to satisfy orthonormality constraints, rendering the TTNS in canonical form,
\begin{equation}
  \sum_{b_{j}^{1}...b_{j}^{Z-1}n_{j}}
  A_{b_{j}^{1}...b_{j}^{Z-1}{b'}_{i}^{\alpha}}^{n_{j}*}
  A_{b_{j}^{1}...b_{j}^{Z-1}b_{i}^{\alpha}}^{n_{j}}
 =\delta_{{b'}_{i}^{\alpha}b_{i}^{\alpha}}.
\end{equation}
As a result, the renormalized basis states $\ket{b_{i}^{\alpha}}$ are  orthonormal. 

The above mathematical properties make a DMRG energy optimization algorithm for TTNS
very similar to that for MPS. Similarly to the DMRG algorithm for MPS, we optimize one site
at a time. The TTNS is expressed in canonical form around site $i$, then the coefficient tensor 
$\bm{\psi}^{n_{i}}$ is optimized, and the sites of the tree are traversed during the sweep.
The  computational challenge is once again how to efficiently compute the representation of the Hamiltonian
in the renormalized bases, namely $\braket{b_{i}^{1}...b_{i}^{Z}n_{i}|\hat{H}|{b'}_{i}^{1}...{b'}_{i}^{Z}{n'}_{i}}$, 
and its action on the coefficient vector $\bm{\psi}^{n_{i}}$, $\hat{H}\psi$. As we discussed for the case
of MPS, for quantum chemistry Hamiltonians the complementary operator approach is most natural.

We rewrite the Hamiltonian in complementary operator form by partitioning into $Z+1$ blocks, A, B, C, D, and so on,
corresponding to the $Z$ branches around site $i$, and site $i$ itself. The Hamiltonian
is re-expressed in terms of operators acting on each of the blocks separately,
\begin{equation}
\begin{split}
  \hat{H}
&=\hat{H}_{A} + \hat{H}_{B} + \hat{H}_{C} + \hat{H}_{D} + ... \\
& \quad
  + \hat{H}_{AB}  + \hat{H}_{AC}  + \hat{H}_{BC} + ... \\
& \quad
  + \hat{H}_{ABC} + \hat{H}_{ACD} + \hat{H}_{BCD} + ... \\
& \quad
  + \hat{H}_{ABCD} + ...
\end{split}
\end{equation}
where,
\begin{widetext}
\begin{equation}
\label{eq:Hdiv2}
  \hat{H}_{AB}
 =\sum_{i\in{A},j\in{B}}
 (\hat{a}_{i}^{A\dag} \hat{S}_{i}^{B}
 +\hat{a}_{j}^{B\dag} \hat{S}_{j}^{A}
 +\hat{a}_{i}^{A\dag} \hat{a}_{i}^{A} \hat{Q}_{ii}^{B})
 +\sum_{i>j\in{A}}
 (\hat{a}_{i}^{A}     \hat{a}_{j}^{A} \hat{P}_{ij}^{B\dag}
 +\hat{a}_{i}^{A\dag} \hat{a}_{j}^{A} \hat{Q}_{ij}^{B})
 + \textrm{adjoint.}
\end{equation}

\begin{equation}
\label{eq:Hdiv3}
\begin{split}
  \hat{H}_{ABC}
&=\sum_{i\in{A},j\in{B}}
  \left(
  \hat{a}_{i}^{A} \hat{a}_{j}^{B} \hat{P}_{ij}^{C\dag} 
 +\hat{a}_{i}^{A\dag} \hat{a}_{j}^{B} \hat{Q}_{ij}^{C\dag} 
  \right)
 +\sum_{i\in{A},j\in{C}}
  \left(
  \hat{a}_{i}^{A} \hat{a}_{j}^{C} \hat{P}_{ij}^{B\dag} 
 +\hat{a}_{i}^{A\dag} \hat{a}_{j}^{C} \hat{Q}_{ij}^{B\dag} 
  \right) \\
&\quad
 +\sum_{i\in{B},j\in{C}}
  \left(
  \hat{a}_{i}^{B} \hat{a}_{j}^{C} \hat{P}_{ij}^{A\dag} 
 +\hat{a}_{i}^{B\dag} \hat{a}_{j}^{C} \hat{Q}_{ij}^{A\dag} 
  \right)
 +\textrm{adjoint.}
\end{split}
\end{equation}

\begin{equation}
\label{eq:Hdiv4}
  \hat{H}_{ABCD}
 =\frac{1}{2}
  \sum_{\substack{i\in{A},j\in{B} \\ k\in{C},l\in{D}}} v_{ijlk}
  \hat{a}_{i}^{A\dag} \hat{a}_{j}^{B\dag} \hat{a}_{k}^{C} \hat{a}_{l}^{D}
 +\textrm{permutation.}
\end{equation}
\end{widetext}

As in the case of MPS, the full matrix representation of $\hat{H}$ is never built explicitly as the
storage requirements would be immense. Instead only $\hat{H}\psi$ is computed
in the Davidson algorithm.
Note that when computing $\hat{H}\psi$
the order of multiplication of the various terms in the complementary operator
decomposition of $\hat{H}$ is important.
For example, the term $\hat{a}_{i}^{A} \hat{a}_{j}^{B} \hat{P}_{ij}^{C\dag} \psi$
in which A and C are large blocks with $\mathcal{O}(k)$ sites and \textit{M} renormalized 
states (indices $b_{1}$ and $b_{2}$ below)
and B is the site being optimized, with 4 states (index $n_{i}$ below) should be computed as 
\begin{widetext}
\begin{equation}
  \sum_{i\in{A},j\in{B}} [\hat{a}_{i}^{A}] \otimes [\hat{a}_{j}^{B}] \otimes [\hat{P}_{ij}^{C\dag}] \boldsymbol{\psi}^{n_{i}}
 =\sum_{i\in{A},b_{1}} [\hat{a}_{i}^{A}]^{b'_{1}b_{1}}
 \left(\sum_{b_{2},n_{i}} \psi_{b_{1}b_{2}...b_{Z}}^{n_{i}}
 \left(\sum_{j\in{B}} [\hat{a}_{j}^{B}]^{n'_{i}n_{i}} [\hat{P}_{ij}^{C\dag}]^{b'_{2}b_{2}}\right)\right).
\end{equation}
\end{widetext}
In the above form,  $\sum_{j\in{B}} [\hat{a}_{j}^{B}] \otimes [\hat{P}_{ij}^{C\dag}]$ is of $\mathcal{O}(16M^{2}k)$ cost,
$[\hat{R}_{i}^{BC\dag}]\psi^{n_{i}}$ is of $\mathcal{O}(64M^{Z+1}k)$ cost,
and finally $\sum_{i\in{A}} [\hat{a}_{i}^{A}] [\hat{R}_{i}^{BC\dag}\psi]$ is of $\mathcal{O}(4M^{Z+1}k)$ cost.

\begin{table}[htb]
\begin{center}
\caption{Complexity of optimal tensor contractions for $\hat{H}\psi$} per site. Multiplying by $\mathcal{O}(k)$ gives complexity per sweep.
\label{tbl:contraction}
\begin{ruledtabular}
\begin{tabular}{ l l }
  terms in $\hat{H}$                                                        & complexity \\
\hline
  $\hat{a}_{i}^{A} \hat{S}_{i}^{B\dag}$                                     & $\mathcal{O}(M^{Z+1}k)$ \\
  $\hat{a}_{i}^{A} \hat{a}_{j}^{A} \hat{P}_{ij}^{B\dag}$                    & $\mathcal{O}(M^{Z+1}k^{2})$ \\
  $\hat{a}_{i}^{A\dag} \hat{a}_{j}^{A} \hat{Q}_{ij}^{B}$                    & $\mathcal{O}(M^{Z+1}k^{2})$ \\
  $\hat{a}_{i}^{A\dag} \hat{a}_{j}^{B\dag} \hat{P}_{ij}^{C}$                & $\mathcal{O}(M^{Z+1}k^{2}+M^{Z+1}k)$ \\
  $\hat{a}_{i}^{A\dag} \hat{a}_{j}^{B} \hat{Q}_{ij}^{C}$                    & $\mathcal{O}(M^{Z+1}k^{2}+M^{Z+1}k)$ \\
  $\hat{a}_{i}^{A\dag} \hat{a}_{j}^{B\dag} \hat{a}_{k}^{C} \hat{a}_{l}^{D}$ & $\mathcal{O}(M^{Z+1}k^{2}+M^{Z+1}k+M^{Z}k^{4})$ \\
\end{tabular}
\end{ruledtabular}
\end{center}
\end{table}


We have carefully optimized the order of tensor contractions for each of the terms in (\ref{eq:Hdiv2})--(\ref{eq:Hdiv4}),
obtaining the complexities  shown in Table~\ref{tbl:contraction}.
For a general $Z$ degree tree,  the total cost
of a single $\hat{H}\psi$ (sigma vector) computation is $\mathcal{O}(M^{Z+1}k^{3}+M^{Z}k^{5})$ per sweep.
To this must be added the
 construction of the matrix representations of the components of $\hat{H}$ in the
different blocks (renormalization steps), which costs $\mathcal{O}(M^{Z+1}k^{3}+M^{Z}k^{5})$ per sweep
 (The renormalization is described explicitly in Appendix A).
In the case of $Z=2$ (MPS) we have already noted that the cost per sweep is lower than in the
general case ($\mathcal{O}(M^{3}k^{3}+M^{2}k^{4})$). This is due to the 
absence of several terms in Table~\ref{tbl:contraction}. The same
is true for the $Z=3$ tree, where the total cost
of a sweep is $\mathcal{O}(M^{4}k^{3}+M^{2}k^{4})$ ($\hat{H} \psi$) and $\mathcal{O}(M^{4} k^{3} + M^{2} k^{5})$ (renormalization).
Because of the special efficiency of the $Z=3$ tree, our later computations focus on this kind of tree.

\begin{figure}
\includegraphics[scale=0.5]{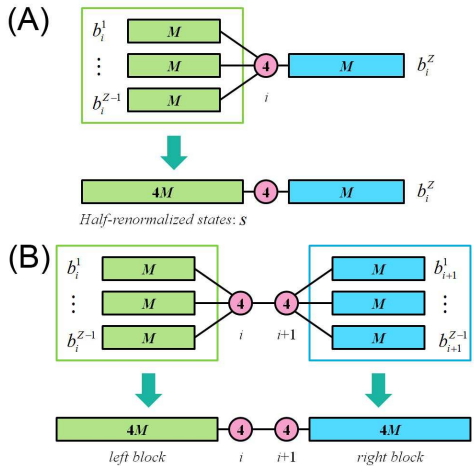}
\caption{Half-Renormalization (HR) algorithm on TTNS.
(A) One-site algorithm: $Z-1$ system blocks $\{\ket{b_{i}^{\alpha}}\}$ are mapped into one system block.
(B) Two-site algorithm: additionally, $Z-1$ environment blocks $\{\ket{b_{i+1}^{\beta}}\}$ are mapped into one environment block.
The half renormalized block contains only $4M$ states for any $Z>2$.}
\label{fig:hr_ttns_algo}
\end{figure}

\subsection{Half-renormalization}

We have found additionally that it is possible to significantly reduce the computational prefactor of a TTNS DMRG calculation
through an additional step we call \textit{half-renormalization} (Fig.~\ref{fig:hr_ttns_algo}).
Half-renormalization involves first constructing an \textit{exact} mapping of the TTNS onto an equivalent $Z=2$ MPS,
then carrying out the $\hat{H}\psi$ operations in this simpler representation. Although the mapping is itself
expensive (and retains the full computational scaling of the TTNS sweep described above), it need
only be carried out once per site, while the $\hat{H} \psi$ operations typically need to be carried out many times
per site during a Davidson diagonalization. To map a TTNS onto an MPS, 
we consider the coefficient tensor at site $i$, $\bm{\psi}_{b^{1}...b^{Z}}^{n_{i}}$. Through an SVD, this tensor can
be exactly decomposed into a rank-3 tensor and a residual tensor, 
\begin{equation}
\begin{split}
  A_{b^{1}...b^{Z}}^{n_{i}\textrm{(TTNS)}}
&=\sum_{ss'}U_{b^{1}...b^{Z-1}s}S_{ss'}V_{s'b^{Z}}^{n_{i}\dag} \\
&=\sum_{s}U_{b^{1}...b^{Z-1}s}\bm{\psi}_{sb^{Z}}^{n_{i}\textrm{(MPS)}}
\end{split}
\end{equation}
where $U$ is the residual tensor and $SV^{\dag}$ is an MPS-like 
pseudo-coefficient tensor $\bm{\psi}_{sb^{Z}}^{n_{i}\textrm{(MPS)}}$.
Note that $S$ is a diagonal matrix with only $4 \times M$ non-zero singular values,
thus $\bm{\psi}_{sb^{Z}}^{n_{i}\textrm{(MPS)}}$ has $O(M^2)$ values, similar
to an MPS coefficient tensor with $M$ renormalized states.
$U_{b^{1}...b^{Z-1}s}$ defines the half-renormalization mapping 
from the states on $Z-1$ branches, $\alpha=1 \ldots Z-1$ of a general tree to 
a \textit{single} set of $4M$ renormalized states $\ket{b_i^s}$ on a single effective branch.
%
These new ``half-renormalized'' states, together with the 
 states on the remaining branch $\alpha=Z$, define the left and right states of an effective MPS, together
with the pseudo-coefficient tensor $\bm{\psi}_{sb^{Z}}^{n_{i}\textrm{(MPS)}}$.
In the case of $Z=3$, the mapping costs  $\mathcal{O}(M^4k^3+M^2k^5)$ per sweep which
leads to the same scaling as the one-site algorithm without half-renormalization.
However, the $\hat{H} \psi$ operations now
carry the MPS cost, namely $\mathcal{O}(M^3 k^3)$ per sweep, leading to a significant savings in computation time.
As the mapping is exact, there is no approximation involved, 
although one drawback is that we typically see slower convergence, as at a given site we are optimize only the pseudo-coefficient
tensor rather than the full coefficient tensor of the TTNS.

\begin{figure}
\includegraphics[scale=0.6]{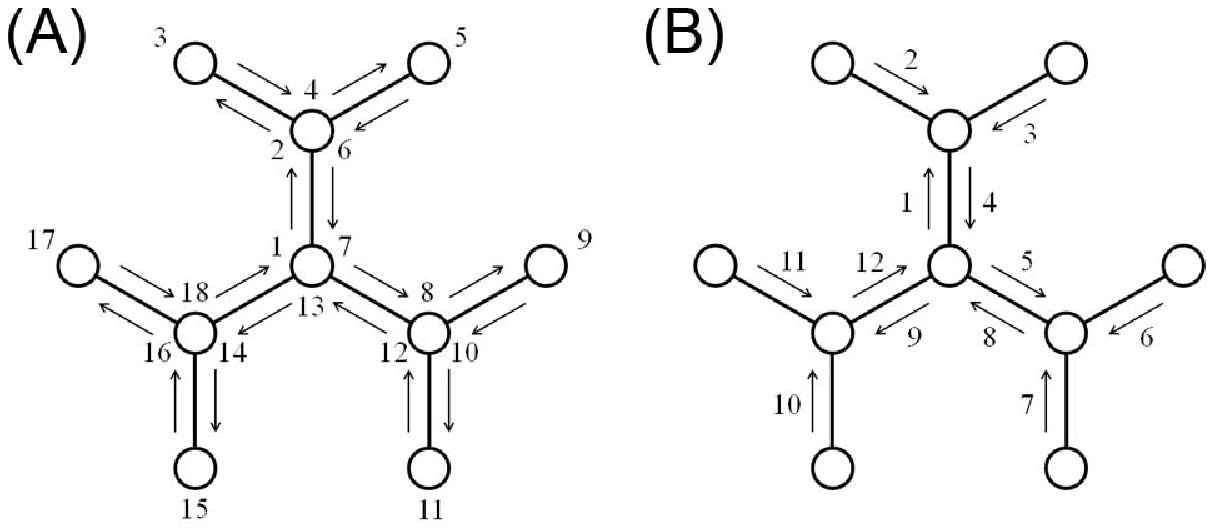}
\caption{Sweep algorithm on a  tree ($Z=3$), by depth-first search with backtracking, where the labels indicate
the order of the search.
(A) One-site algorithm: starting from the center site, 1 sweep contains 18 one-site optimization steps, and
(B) Two-site algorithm: starting from the center site and one adjacent site, 1 sweep contains 12 two-site optimization steps.}
\label{fig:ttns_sweep_algo}
\end{figure}

The full DMRG sweep on a tree is carried out using depth-first search with backtracking.
Applying this to a $Z=2$ tree, as in an MPS, recovers the usual DMRG sweep algorithm.
Figure~\ref{fig:ttns_sweep_algo}(A) shows the details of the one-site sweep algorithm on TTNS.

\begin{description}
\item[(1)] Define any site as the root (depth = 0). Construct the
           TTNS in canonical form at the site, construct the
           renormalized states and operators by contracting from the leaves to the root site.
\item[(2)] Optimize the wavefunction at site $i$ with Davidson diagonalization.
           Renormalize to construct the canonical form at a neighbouring site.
\item[(3)] Continue (2) by carrying out a depth-first search with backtracking to optimize the entire tree.
\item[(4)] Continue (2)-(3) until the energy is converged to a target accuracy.
\end{description}

\subsection{Two-site TTNS algorithm}

\begin{figure}
\includegraphics[scale=0.5]{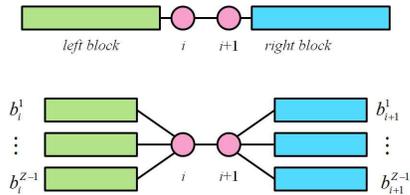}
\caption{Block structure of the two-site algorithm.
Top panel shows the two-site MPS which contains one system block, system site, environment site, and one environment block.  Bottom panel shows the two-site TTNS which contains $Z-1$ system blocks, system site, environment site, and $Z-1$ environment blocks.}
\label{fig:twosite_graph_rep}
\end{figure}

Although our discussion of MPS focused for simplicity on the one-site MPS and the corresponding one-site DMRG algorithm,
it is well known from practical experience that one-site DMRG calculations tend to get stuck in local minima
and suffer from poor convergence characteristics. Thus, the two-site MPS and DMRG algorithm are more commonly used.\cite{chan2002,schollwock2005,white2005}
In the case of MPS, the two-site MPS wavefunction is obtained by modifying the coefficient tensor to span two sites,
\begin{align}
  \ket{\Psi}
 =\sum_{l_{i-1}n_{i}n_{i+1}r_{i+1}} \psi_{l_{i-1}r_{i+1}}^{n_{i}n_{i+1}} \ket{l_{i-1}n_{i}n_{i+1}r_{i+1}}.
\end{align}
A $Z$-branch two-site TTNS is modified in a similar way
\begin{align}
  \sum_{\substack{n_{i}n_{i+1}\\\mathbf{b}_{i}\mathbf{b}_{i+1}}}
  \psi_{b_{i}^{1}...b_{i}^{Z-1}b_{i+1}^{1}...b_{i+1}^{Z-1}}^{n_{i}n_{i+1}} \ket{b_{i}^{1}...b_{i}^{Z-1}n_{i}n_{i+1}b_{i+1}^{1}...b_{i+1}^{Z-1}}
\end{align}
An important difference however, between a general two-site tree and two-site MPS is that whereas around sites $i, i+1$ in
an MPS we can define left and right renormalized basis, just as for an one-site MPS, in the case of a TTNS, there are $2Z-2$ branches
around sites $i, i+1$ (Fig.~\ref{fig:twosite_graph_rep}), rather than the $Z$ branches around a single-site. 
Consequently, the naive cost of the $\hat{H}\psi$ operations in a two-site TTNS DMRG sweep using complementary operators is much higher than that for an one-site TTNS sweep, with  a prohibitive cost of $\mathcal{O}(M^{2Z-1}k^{3}+M^{2Z-2}k^{5})$. However, by employing
two sets of half-renormalization steps, we can map the two-site TTNS onto a two-site MPS, reducing the cost of the $\hat{H}\psi$ operations to only $\mathcal{O}(M^3 k^3)$ per sweep with an additional half-renormalization cost of
 $\mathcal{O}(M^{Z+1}k^{3}+M^{Z}k^{5})$ per sweep for general $Z$ degree trees, or $\mathcal{O}(M^4k^3+M^2k^5)$ per sweep in the case of $Z=3$. This is the same cost as an one-site TTNS sweep with half-renormalization,
and thus the two-site TTNS sweep becomes practical. We compare the cost and convergence characteristics of the one-site
and two-site TTNS sweeps with half-renormalization in our later calculations.

\subsection{Tree Shape and Site Ordering}

\begin{figure}[htb]
\includegraphics[scale=0.5]{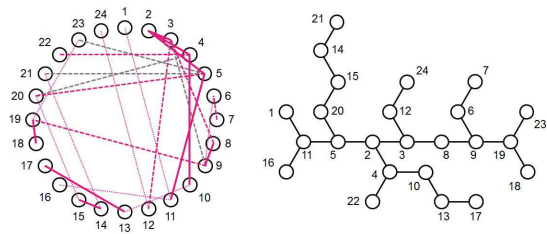}
\caption{Approximate degree-fixed minimum spanning tree (MST, $Z = 3$) for 24 orbitals from a RHF/cc-pVDZ calculation of
the  water molecule.
Left panel shows a representation of the exchange integral $K_{ij}$,
where solid lines denote $K_{ij} \geq 0.10$, dashed lines denote $K_{ij} \geq 0.07$, and dotted lines denote $K_{ij} < 0.07$.
Colored lines are connected lines in the right panel and gray lines are ignored interactions. For $K_{ij} < 0.07$, only connected lines are shown.
Labels in the right panel indicate the MOs (as indexed by  energy).}
\label{fig:h2o_tree_mst}
\end{figure}

\begin{figure}[htb]
\includegraphics[scale=0.6]{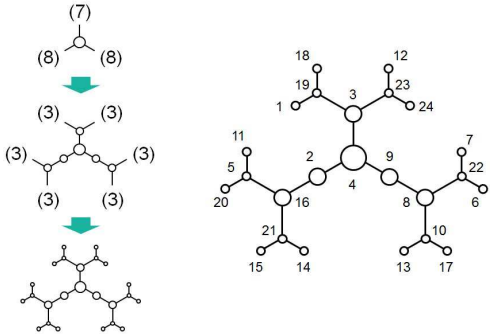}
\caption{Minimum-entangled tree (MET, $Z = 3$) for 24 orbitals from a RHF/cc-pVDZ calculation of the water molecule.
Left panel shows the stepwise construction of MET. First, 24 sites are divided into 1 center site and 3 branches containing 7, 8, and 8 sites.
Second, 7 sites are divided into 1 site and 2 branches containing 3 sites for each, and 8 sites are divided into 2 sites and 2 branches containing 3 sites for each.
Finally, we used a genetic algorithm to map the orbitals to the tree sites as shown in right panel.
Labels in right panel indicate MO indices (as indexed by energy).}
\label{fig:h2o_tree_met}
\end{figure}

\begin{table*}[htb]
\caption{The ground state energies of water molecule calculated with two different trees.}
\label{tbl:results_tree_shape}
\begin{ruledtabular}
\begin{tabular}{ c c c c c }
 \multirow{2}{*}{\textit{M}} & \multicolumn{2}{c}{MST}              & \multicolumn{2}{c}{MET}              \\ \cline{2-5}
                             & \textit{E} / a.u. & CPU time / sec. sweep & \textit{E} / a.u. & CPU time / sec. sweep \\
\hline
                 100         & -76.242805      &  307.9           & -76.242242      &  220.6           \\
                 200         & -76.243652      & 1209.9           & -76.243491      &  826.5           \\
\end{tabular}
\end{ruledtabular}
\end{table*}

When carrying out a quantum chemistry DMRG calculation using an MPS, it is necessary to choose a
mapping of the sites to the 1D lattice. Generally, this should be done to minimize 
entanglement between distant sites on the lattice,\cite{legeza2003b,white2006,reiher2011}
but computing the entanglement and carrying out an exact minimization are costly procedures.
In practice, an approximate proxy for the entanglement
between orbitals is constructed, and it is approximately minimized.\cite{chan2002} One example of such a
proxy which has been used in prior DMRG studies is a weighted exchange integral $K_{ij}D_{ij}^n$, where
$K_{ij}$ is the exchange integral between orbitals $i$ and $j$, $D_{ij}$ is the separation
on the lattice, and $n$ is an adjustable parameter.



In the case of ordering for trees, we need to consider not only the mapping
of the orbitals onto a given tree, but also the shape of the tree as well, even
if we restrict ourselves to trees of a fixed maximum degree.
Since the computational cost of a sweep also depends on the tree shape (changing
the prefactor associated with the cost of the sweep) the tree which gives the lowest energy
for a given $M$ is  not necessarily the most efficient tree to use in practice.

To illustrate these issues we consider the  water molecule in a cc-pVDZ basis set, which
has 24 orbitals. We have considered two shapes of trees:
a degree-fixed Minimum-Spanning Tree (MST)
 shown in Figure~\ref{fig:h2o_tree_mst}, and a Minimum-Entangled Tree (MET).
The MET is defined as the shape of tree, for a given degree, where the number of renormalized states required to achieve an {\it exact} calculation
is minimized. Its construction is shown in Figure~\ref{fig:h2o_tree_met}.
Because of its balanced nature, the MET also minimizes the prefactor of the cost of the TTNS calculation 
for a given $M$.

For each of these trees, we mapped the orbitals onto the sites by a
genetic algorithm\cite{sivanandam2010} that minimized the cost function $\sum_{ij} K_{ij} D_{ij}^2$,
where $D_{ij}$ is the counting distance between the sites $i$ and $j$ in the tree.

The ground state energies computed with the two different trees and two different \textit{M} (100 and 200)
are summarized in Table~\ref{tbl:results_tree_shape}.
From these calculations, we see that the MST gives a slightly better energy than that the MET,
but the MST sweep is 50\% slower than the MET sweep for the same \textit{M}, due to the
unbalanced nature of the MST.
For larger $M$, because the MET minimizes the number of renormalized states for the exact calculation, we expect
it to eventually give a lower energy for a given $M$ than the corresponding MST.
Although these results are system dependent, they
 indicate the importance of tree shape
in determining the cost of practical TTNS calculations.
We adopt the MET in our calculations unless another tree shape is trivially indicated
by the molecular structure, e.g. in a tree shaped molecule such as a dendrimer.

\section{Illustrative Calculations}

\begin{figure}
\includegraphics[scale=0.6]{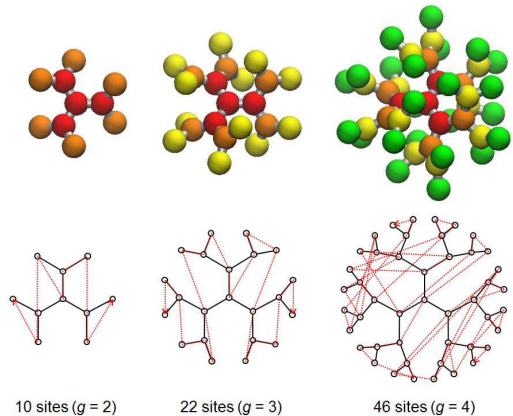}
\caption{Hydrogen atoms on Cayley-trees (\textit{Z} = 3, \textit{g} = 2 (10 sites), 3 (22 sites), and 4 (46 sites)).
         Top panel shows the actual structures of the hydrogen trees,
         in which red atoms are the core region, orange atoms are for \textit{g} = 2,
         yellow atoms are for \textit{g} = 3, and green atoms are for \textit{g} = 4.
         The distance between adjacent hydrogens is 2.0 Bohr and torsional angle between different generations is 30 degrees.
         Bottom panel shows corresponding Cayley-tree diagrams. The red-dotted line denotes the site ordering used in the MPS.}
\label{fig:htree_image}
\end{figure}

\begin{figure}
\includegraphics[scale=0.6]{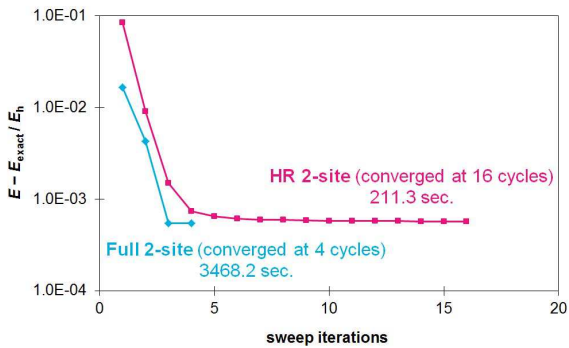}
\caption{Energy convergence and total CPU time computed with normal two-site TTNS (denoted Full)
         and half-renormalized two-site TTNS (denoted HR-TTNS).}
\label{fig:htree10_hr_result}
\end{figure}

\begin{figure*}
\includegraphics[scale=0.6]{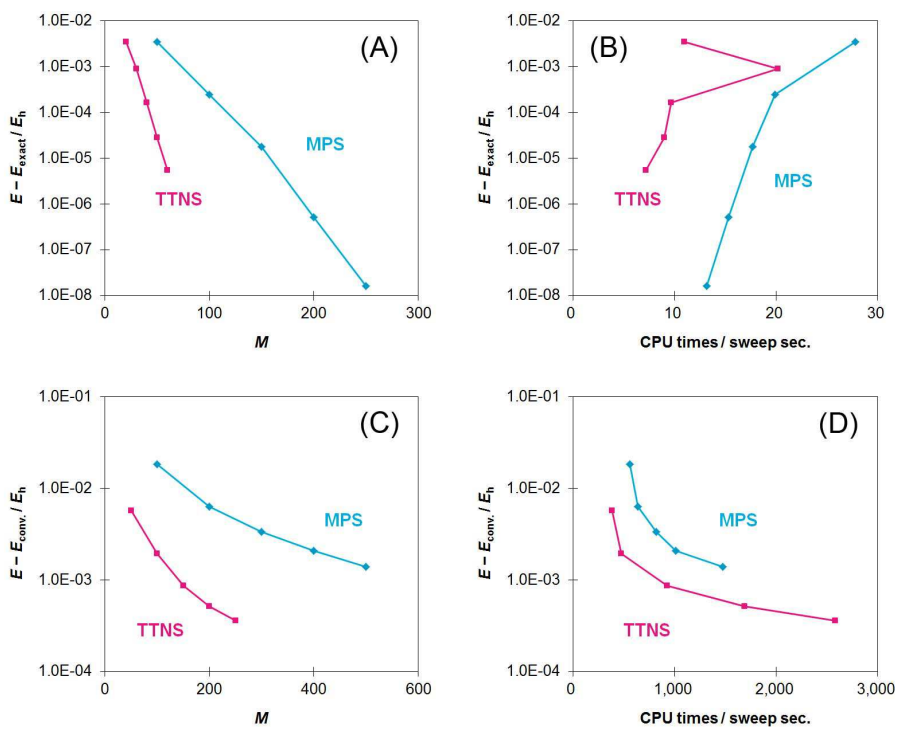}
\caption{Energy convergence of hydrogen trees
         plotted with respect to the number of renormalized states \textit{M} and CPU time (sec.) per one sweep;
         (A) and (B) show energy versus \textit{M} and CPU time per sweep, respectively, for 10 sites,
         and (C) and (D) show energy versus \textit{M} and CPU time per sweep, respectively, for 22 sites.
         All calculations are for the  triplet ground state energy in an orthogonalized STO-3G basis.}
\label{fig:htree_results}
\end{figure*}

\begin{figure*}
\includegraphics[scale=0.6]{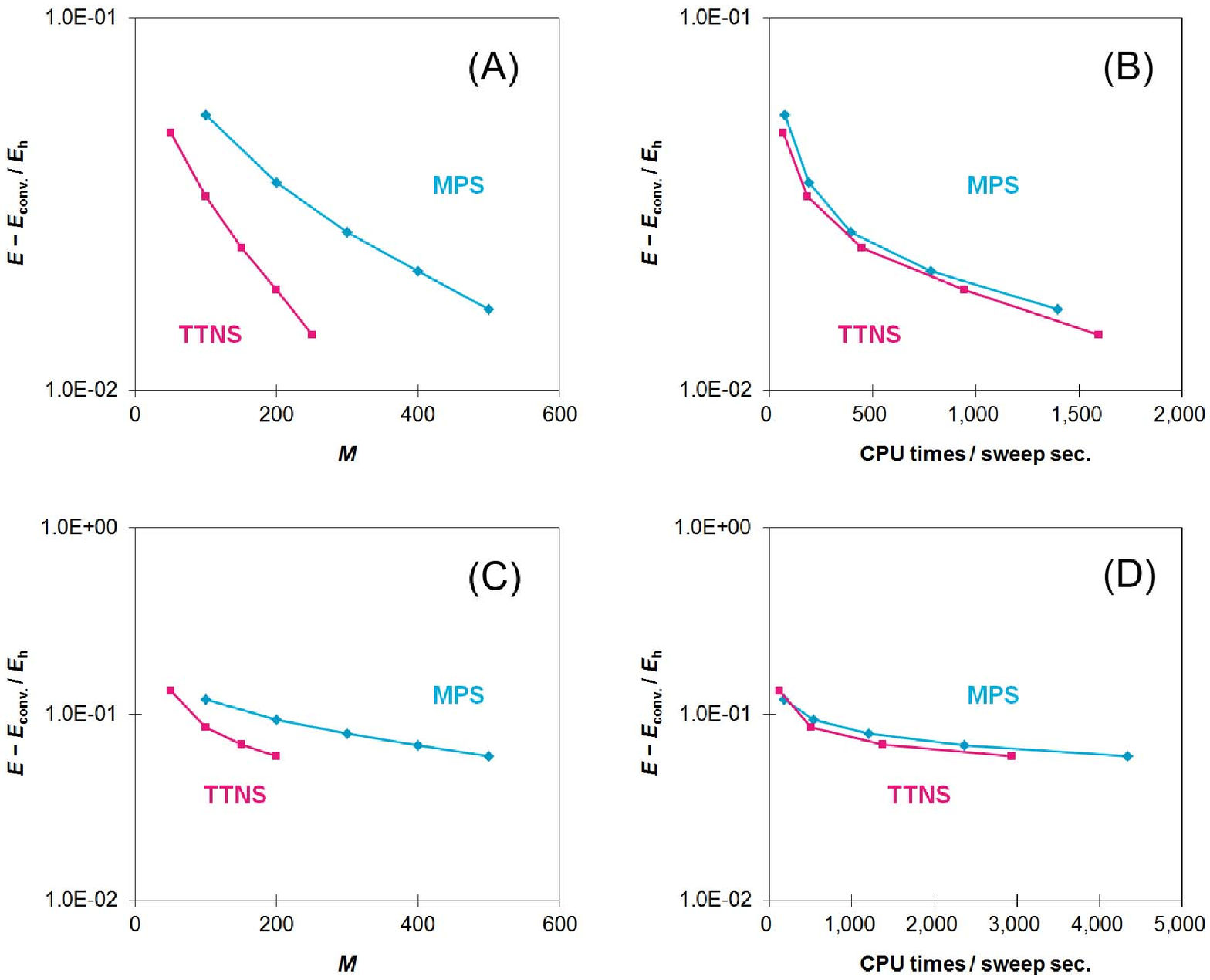}
\caption{Energy convergence of hydrogen chains with CMOs
         plotted by number of renormalized states \textit{M} and CPU time (sec.) per one sweep;
         (A) and (B) showed those of \textit{M} and CPU time per sweep, respectively, for 20 sites,
         and (C) and (D) showed those of \textit{M} and CPU time per sweep, respectively, for 30 sites.
         Singlet ground state energy was computed with STO-3G basis sets.}
\label{fig:hchain_results}
\end{figure*}

We now present several benchmark applications to molecular systems to understand the performance of TTNS and to
 compare with that of MPS.
For the MPS calculations, we used our TTNS code with $Z=2$. This was to allow a fair comparison of timings using
the same implementation.

To start, we consider an idealized system: hydrogen atoms on Cayley-trees in an (orthogonalized) minimal STO-3G basis. This is a model system
where TTNS are expected to work very well.
To avoid nearly overlapping  hydrogens which would arise in a planar geometry,
the tree structures are taken to be slightly twisted as shown in Figure~\ref{fig:htree_image}. All geometries
are provided in the supplementary information.\cite{supinfo}

We first illustrate the importance of the half-renormalization algorithm by comparing the normal
two-site TTNS algorithm and the HR-TTNS two-site  algorithm on a small 10 site hydrogen tree.
The energy convergence along the sweeps and total CPU time are shown in Figure~\ref{fig:htree10_hr_result}.
Though the convergence per sweep is slower in the HR-TTNS algorithm, the total CPU time is much smaller than in the normal two-site TTNS 
algorithm due to the much less expensive $\hat{H}\psi$ operation. 
Thus, the remaining calculations have been done using the HR-TTNS algorithm, which we henceforth shorten to TTNS.

Next, we compare the performance of the MPS and TTNS on three hydrogen trees of different sizes: 10 sites, 22 sites, and 46 sites.
We focus on two aspects: energy convergence with \textit{M}, and CPU time per sweep.
In general, the energy convergence of the TTNS is much faster than the MPS, both  with increasing \textit{M} and 
both as a function of CPU time, as seen in Figure~\ref{fig:htree_results} for the 22 site tree.
(In the 10 site tree, the strange behaviour of the CPU time, where larger \textit{M} required smaller CPU time,
is an artifact of our Davidson diagonalization implementation, which required a large number of iterations
when $M$ was very small).
In the 46 site tree, we only carried out calculations with small \textit{M}  because of the large computational cost.
For the CPU time per sweep, we found that a calculation with MPS with $M=200$ (6182 sec. / sweep), and TTNS with $M=100$ (6839 sec. / sweep) were comparable.
However, the corresponding energies were computed to be $-22.705014 E_h$ and $-22.958219 E_h$, respectively and thus
 the TTNS gave a much better energy than the MPS at a comparable computational cost.
These results demonstrate that  TTNS are more cost effective than MPS when the system is tree-shaped.
Moreover, the relative benefits of the TTNS increase as the tree  size increases.

We now consider a model system designed to mimic more difficult molecular structures for MPS and TTNS. In the previous 
tree structures
we used a local atomic basis. This, together with the underlying tree connectivity of the molecule, allowed
the TTNS to completely exploit the local nature of correlation in the system. Though the system was not linear,  the MPS
still benefitted from the local basis in the calculations, as not every orbital was correlated with every other. 
(For similar reasons, in DMRG calculations localized molecular orbitals are often used to minimize 
long-range entanglement). However, in some situations, it may be unavoidable 
for an MPS or TTNS to describe some long-range entanglement, either due to the underlying physics,
or due to a poor mapping of the molecular structure onto a chain or tree. To mimic this situation
we carry out MPS and TTNS calculations in a {\it canonical} molecular orbital basis. Because
these orbitals are delocalized,  every orbital can be considered to 
interact with every other.
We choose as our model system a set of hydrogen chains in a minimal STO-3G basis. 
Although these are chain molecules, the use of canonical molecular orbitals means that this
is not an ideal system for the MPS.

Figure~\ref{fig:hchain_results} shows the energy convergence as a function of \textit{M} and CPU time per sweep.
Both MPS and TTNS exhibit very slow energy convergence as a function of \textit{M} due
to the use of canonical molecular orbitals, but it is clear that the convergence with \textit{M} is always 
better in TTNS than in MPS by a factor of 2 or more. As a function of CPU time, we find that the decreased
$M$ means that TTNS performs better in comparison with MPS, although the improvement is slight due
to the higher scaling of TTNS with $M$.

These results provide some promise that in general molecules without a tree or chain structure,
TTNS can perform better than MPS, although this involves a delicate balance between the
decreased number of renormalized states used by the TTNS, and the higher computational scaling with $M$.  
To examine this in a more realistic setting, we now consider calculations on two benchmark diatomics, the nitrogen dimer and the chromium dimer.

\begin{figure}
\includegraphics[scale=0.6]{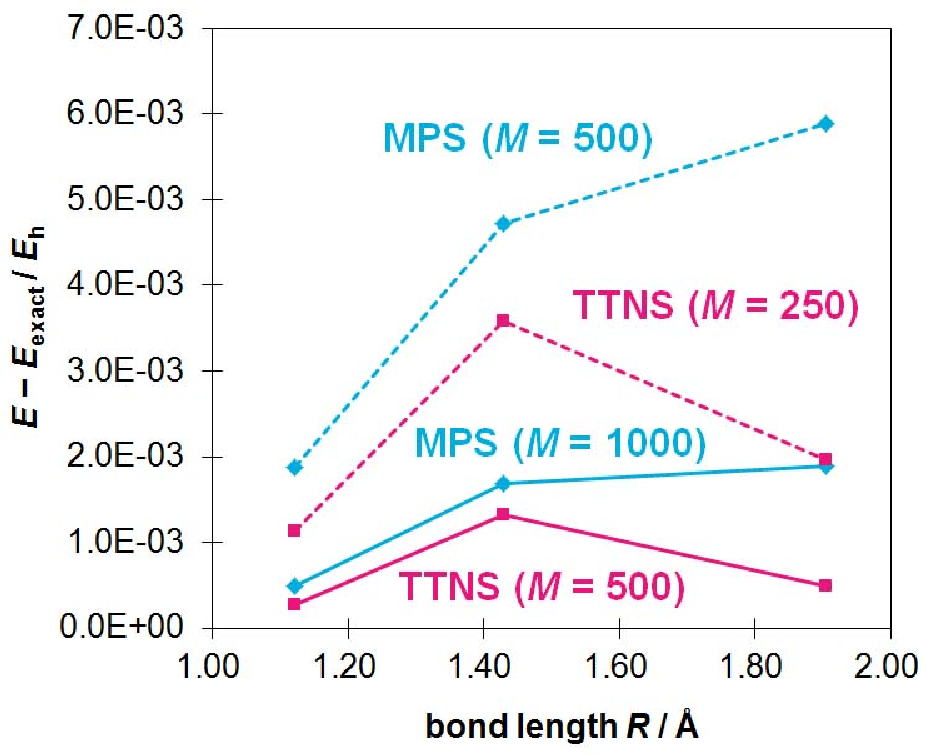}
\caption{Energy errors for bond dissociation of nitrogen dimer using (10e, 26o) active space and cc-pVDZ basis sets.
         Reference energies were from previous full-CI work in Ref.~\onlinecite{larsen2000}.}
\label{fig:nitrogen_results}
\end{figure}

\begin{table}
\caption{CPU times per sweep in sec. for MPS and TTNS calculations for the nitrogen dimer using (10e, 26o) active space and cc-pVDZ basis sets.}
\label{tbl:nitrogen_results}
\begin{ruledtabular}
\begin{tabular}{ c c c c c c }
 \multirow{2}{*}{\textit{R} / \text{\AA}} & \multicolumn{2}{c}{MPS}              & & \multicolumn{2}{c}{TTNS}            \\ \cline{2-3}\cline{5-6}
                                          & \textit{M} = 500 & \textit{M} = 1000 & & \textit{M} = 250 & \textit{M} = 500 \\
\hline
                1.1208                    &              706 &              2998 & &              794 &             5088 \\
                1.4288                    &             1027 &              4684 & &             1337 &             7510 \\
                1.9050                    &             1155 &              5053 & &              992 &             5945 \\
\end{tabular}
\end{ruledtabular}
\end{table}

\begin{figure*}
\includegraphics[scale=0.6]{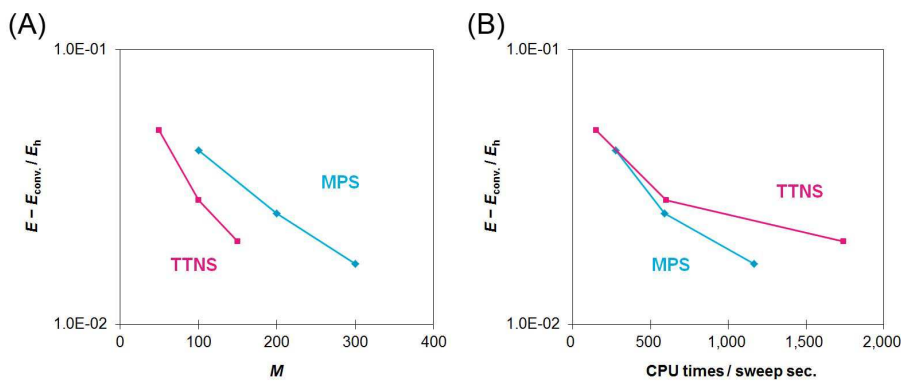}
\caption{Energy convergence of the chromium dimer at 1.5 \text{\AA} using a (24e, 30o) active space and Ahlrichs' SVP basis set.
         (A) Plotted as a function of \textit{M} and (B) as a function of CPU time per sweep.
         Reference energies were from previous DMRG work in Ref.~\onlinecite{sharma2012}.}
\label{fig:chromium_results}
\end{figure*}

\begin{figure}
\includegraphics[scale=0.4]{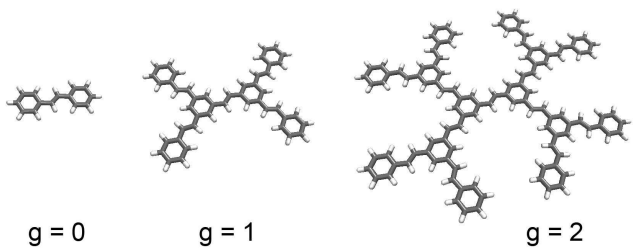}
\caption{Structures of stilbenoid dendrimers \textit{g} = 0, 1, and 2.
         Geometries were optimized at the B3LYP/cc-pVDZ level of theory.}
\label{fig:stilbenoid_dendrimers}
\end{figure}

\begin{figure*}
\includegraphics[scale=0.6]{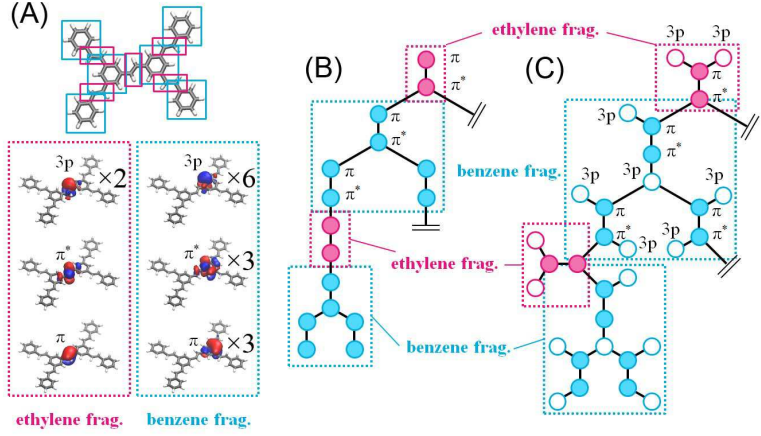}
\caption{Tree graphs for TTNS calculation of stilbenoid dendrimer (\textit{g} = 1).
         (A) Selected molecular orbitals of the localized ethylene and benzene fragments,
         (B) Tree graph of the single-valence (STO-3G) calculation, and
         (C) tree graph of the double-valence (6-31G) calculation
         where the filled circles represent local $\pi$ and $\pi^{*}$ orbitals and empty circles represent local $3p$ orbitals.
         Note that identical orbitals are omitted in the tree graphs.} 
\label{fig:b2g1_orbital_tree}
\end{figure*}

\begin{figure*}
\includegraphics[scale=0.6]{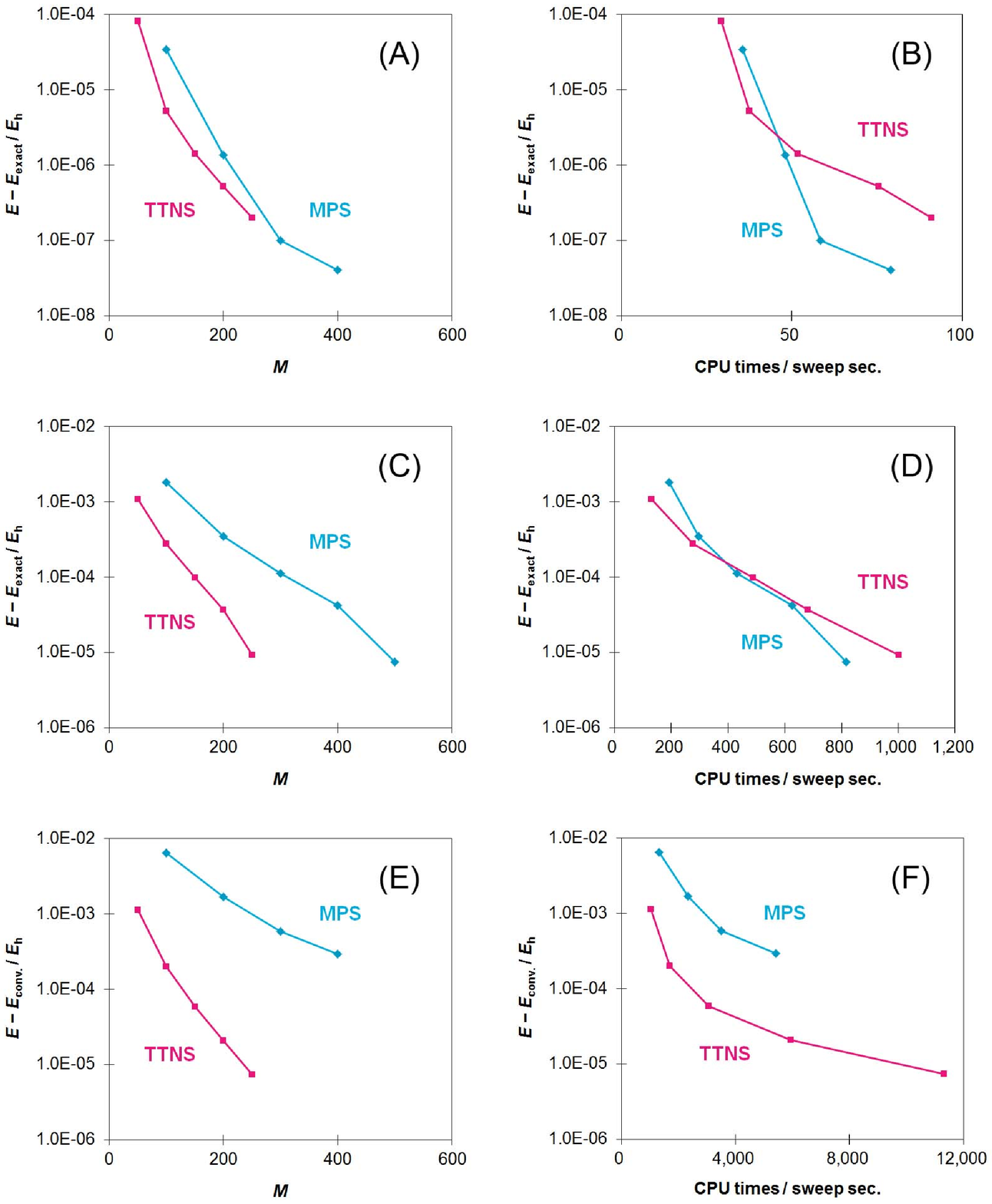}
\caption{Energy convergence of stilbenoid dendrimers \textit{g} = 0 and \textit{g} = 1
         plotted by number of renormalized states $M$ and CPU time (sec.) per sweep;
         (A) and (B) shows energy versus $M$ and CPU time per sweep, respectively, for \textit{g} = 0, single valence (14e, 14o),
         (C) and (D) shows energy versus $M$ and CPU time per sweep, respectively, for \textit{g} = 0, double valence (14e, 28o),
         (E) and (F) shows energy versus $M$ and CPU time per sweep, respectively for \textit{g} = 1, single valence (46e, 46o).
         All energies are for the singlet ground-state energy.}

\label{fig:stilbenoid_results}
\end{figure*}

The bond dissociation curve of nitrogen dimer is often used as a good benchmark to evaluate whether a method can describe strong electron correlation correctly.
We evaluated ground state energies at three points on the bond dissociation of nitrogen dimer, 1.1208($R_{e}$), 1.4288, and 1.9050 \text{\AA}.
A frozen core active space (10e, 26) with a cc-pVDZ basis set was employed as used in previous DMRG calculations\cite{chan2002} and full-CI calculations\cite{larsen2000} (see supplementary information for the tree graphs and the site orderings\cite{supinfo}).
Figure~\ref{fig:nitrogen_results} shows the energy errors from full-CI results\cite{larsen2000} as a function of bond length and
CPU times per sweep are summarized in Table~\ref{tbl:nitrogen_results}.
In TTNS, half the \textit{M} can be used at \textit{R} = 1.1208 and 1.4288 \text{\AA} and a quarter the \textit{M} can be used at \textit{R} = 1.9050 \text{\AA} as compared with MPS. TTNS with half the \textit{M} of the corresponding MPS required approximately the same CPU time per sweep.
This indicates that the TTNS gives comparable and/or slightly better performance at short bond-lengths but much better performance at long bond-lengths compared to the MPS in this molecule.
We conclude that the TTNS works better in the case of the nitrogen dimer than the MPS.

Since the chromium dimer has an unusual multiple bond, its ground state is very complicated and difficult
to describe by conventional methods.
Recently, DMRG calculations of the chromium dimer have been carried out for relatively large active spaces.\cite{kura2009,sharma2012}
We performed MPS and TTNS calculation at 1.5 \text{\AA} using the same basis sets and the same active space (24e, 30o)
as in earlier DMRG calculations \cite{sharma2012} (see supplementary information for the tree graph and the site ordering\cite{supinfo}).
Figure~\ref{fig:chromium_results} shows the energy convergence as a function of \textit{M} and CPU time per sweep.
Since the previous DMRG calculation employed very large \textit{M} (up to 10000), the energy reported here is far from convergence because we only employed \textit{M} up to 150 in the TTNS.
Nonetheless compared to our MPS benchmarks, we find that smaller \textit{M} can be used in the TTNS.
In comparing CPU time per sweep, however, we see that MPS gives much better performance than the TTNS in this molecule.
Thus, we conclude that the MPS works better in the case of the chromium dimer than the TTNS.

These two benchmark calculations on the nitrogen dimer and chromium dimer show that the performance of the MPS and TTNS 
in {\it general} molecules depends sensitively on the electronic structure and the nature of the
quantum entanglement of the molecule, thus their relative merits must be determined on a molecule
by molecule basis.

Finally, to demonstrate the power of TTNS, we consider a TTNS calculation on  more realistic
tree-shaped molecules.
Stilbenoid dendrimers are prototypical $\pi$-conjugated dendrimers, with each unit being a stilbene fragment.
These dendrimers are potentially attractive for chemistry because
photo-induced electron-transfer or exciton-transfer may proceed from the leaves to the core,
mimicking a biological antenna system (we note that dendrimer systems have also been previously studied using
semi-empirical DMRG \cite{sierra2002,ramasesha2008}).
Although it is the excited states and dynamical properties of these systems that are of primary interest, here we
focus on the ground-state energy for benchmarking and reserve the study of excited states
to future work. 

We consider three different sizes of stilbenoid dendrimers denoted by the generation g as shown in Figure~\ref{fig:stilbenoid_dendrimers}.
We carried out $\pi$-full valence MPS and TTNS calculations with STO-3G and 6-31G basis sets.
Single valence calculations were performed for \textit{g} = 0, 1, and 2 with (14e, 14o), (46e, 46o), and (110e, 110o) active-spaces, respectively, and double valence calculations were performed for \textit{g} = 0 and 1 with (14e, 28o) and (46e, 92o) active-spaces respectively. 
To construct the tree graph, $\pi$-orbitals computed from a RHF calculation were localized for occupied and unoccupied spaces separately.
Localized MOs were grouped for each ethylene and benzene fragment, and were ordered on the tree according to 
the underlying dendritic structure, as shown in Figure~\ref{fig:b2g1_orbital_tree}.
The orderings within each fragment were determined to put strongly interacting pairs (evaluated by $K_{ij}$) on neighboring sites.

For the \textit{g} = 0 dendrimer (stilbene), the MPS and TTNS gave similar energy convergence as a function of $M$ in the single-valence (STO-3G) 
calculation, and the MPS gave somewhat better performance with respect to CPU time per sweep (see Figure~\ref{fig:stilbenoid_results}A and \ref{fig:stilbenoid_results}B). 
In the double-valence (6-31G) calculation, the TTNS gave an improved energy convergence with $M$ compared to MPS,
and the MPS and TTNS gave almost the same performance with respect to CPU time per sweep (see Figure~\ref{fig:stilbenoid_results}C and \ref{fig:stilbenoid_results}D). The competitive performance of MPS relative to TTNS in the \textit{g} = 0 dendrimer
 reflects the very small size of the system, which is almost linear in nature and thus nearly ideal for MPS.

For the larger \textit{g} = 1 dendrimer, in the single-valence space the TTNS required a quarter the $M$ to obtain  the same energy accuracy as the MPS.
The computational cost of the TTNS for this accuracy was also much lower than that of the 
MPS
(see Figure~\ref{fig:stilbenoid_results}E and \ref{fig:stilbenoid_results}F).
The double-valence calculations were very expensive, hence we only performed  calculations with $M=100$ for MPS and with $M=50$ for TTNS. Although the resulting energies were not converged to chemical accuracy,  
 the correlation energies ($E - E_{\rm{HF}}$)
of the MPS,  $-0.644736 E_h$, and the TTNS, $-0.678678 E_h$, indicate that the TTNS is once again much more accurate.

Finally, for the \textit{g} = 2 dendrimer, even the single-valence active space calculation was quite expensive with
our implementation, consequently, we only 
performed single-valence calculations using $M=100$ for MPS and $M=50$ for TTNS. These calculations had comparable timings.
The computed correlation energies were $-1.943651 E_h$ and $-2.021378 E_h$ with MPS and TTNS, respectively, indicating once
again that the TTNS works much better than the MPS in this large dendritic molecule. 
Although we could not reach fully converged energies  in this work, the ability to even
approximately target such systems with TTNS demonstrates the promise of the technique for complex systems.

\section{Conclusions}

In this work we investigated tree tensor network states (TTNS) for quantum chemistry.
We formulated an efficient tree tensor network algorithm that is
analogous to the density matrix renormalization group (DMRG) algorithm
in quantum chemistry for matrix product states (MPS). We introduced the additional step of half-renormalization 
that greatly reduced the computation cost.
We found that our TTNS calculations were competitive with MPS and DMRG
calculations in general molecules, requiring  significantly fewer renormalized states
for the same accuracy, although this did not always translate into a savings in computational time. In tree
like molecules,  TTNS were clearly superior to MPS requiring both
fewer renormalized states and less time to reach the same accuracy. 
This bodes well for the application of TTNS to study a wide
class of interesting optically active systems based on dendritic structures, as illustrated in our calculations
on stilbenoid dendrimers.

Tree tensor network states are one of the simplest generalizations of the matrix product states,
because the tree network structure has no cycles. More complex tensor networks which
describe even more general entanglement require the treatment of cycles.
The improvements observed with tree tensor networks here suggests that 
investigating these more complex classes will be fruitful, particularly
to describe  quantum chemistry of larger systems, where the molecules or materials
acquire an extended two-dimensional and three-dimensional structure.

\section{Acknowledgements}

This work was supported by the National Science Foundation (NSF) through Grant No. NSF-OCI-1148287 and NSF-CHE-1213933.

\section{Appendix A: Optimal Tensor Contraction for TTNS Renormalization}

During the TTNS renormalization procedure, it is necessary to construct matrix representations
of operators act on ${b_{i}^{1}...b_{i}^{Z-1}n_{i}}=\ket{n_{i}}\ket{b_{i}^{Z-1}}...\ket{b_{i}^{1}}$.
Because of the quartic terms in the Hamiltonian, this can involve 
 products of up to four operators. The corresponding product formulae follow.

The matrix representation of an operator $\hat{O}_{\alpha}$ which acts only on renormalized states in block $\alpha$ is given by,
\begin{equation}
\label{eq:Ops1}
\begin{split}
& [\hat{O}_{\alpha}]^{{b'}_{i}^{1}...{b'}_{i}^{Z-1}{n'}_{i}{b}_{i}^{1}...{b}_{i}^{Z-1}{n}_{i}} \\
&=\bra{{b'}_{i}^{1}}...\bra{{b'}_{i}^{Z-1}}\bra{{n'}_{i}}
  \hat{O}_{\alpha}
  \ket{{n}_{i}}\ket{{b}_{i}^{Z-1}}...\ket{{b}_{i}^{1}} \\
&=[\hat{O}_{\alpha}]^{{b'}_{i}^{\alpha}{b }_{i}^{\alpha}}
  \times \hat{P}(\bra{{b'}_{i}^{\alpha+1}}...\bra{{b'}_{i}^{Z-1}}\bra{{n'}_{i}},\hat{O}_{\alpha})
\end{split}
\end{equation}
where $[\hat{O}_{\alpha}]^{{b'}_{i}^{\alpha}{b }_{i}^{\alpha}}$ is the matrix representation of $\hat{O}_{\alpha}$ spanned by states $\ket{{b }_{i}^{\alpha}}$
and $\hat{P}$ is a parity operator which gives $+1$ or $-1$ depending on the particle numbers of $\bra{{b'}_{i}^{\alpha+1}...{b'}_{i}^{Z-1}{n'}_{i}}$ and $\hat{O}_{\alpha}$.
Similarly, those of operators
$\hat{O}_{\alpha} \hat{O}_{\beta }$,
$\hat{O}_{\alpha} \hat{O}_{\beta } \hat{O}_{\gamma}$, and
$\hat{O}_{\alpha} \hat{O}_{\beta } \hat{O}_{\gamma} \hat{O}_{\delta}$
are computed as follows,
\begin{equation}
\label{eq:Ops2}
\begin{split}
& [\hat{O}_{\alpha} \hat{O}_{\beta }]^{{b'}_{i}^{1}...{b'}_{i}^{Z-1}{n'}_{i}{b}_{i}^{1}...{b}_{i}^{Z-1}{n}_{i}} \\
&=[\hat{O}_{\alpha}]^{{b'}_{i}^{\alpha}{b }_{i}^{\alpha}}
  [\hat{O}_{\beta }]^{{b'}_{i}^{\beta }{b }_{i}^{\beta }} \\
& \quad \times \hat{P}(\bra{{b'}_{i}^{\alpha+1}}...\bra{{b'}_{i}^{Z-1}}\bra{{n'}_{i}},\hat{O}_{\alpha}) \\
& \quad \times \hat{P}(\bra{{b'}_{i}^{\beta +1}}...\bra{{b'}_{i}^{Z-1}}\bra{{n'}_{i}},\hat{O}_{\beta }),
\end{split}
\end{equation}

\begin{equation}
\label{eq:Ops3}
\begin{split}
& [\hat{O}_{\alpha} \hat{O}_{\beta } \hat{O}_{\gamma}]^{{b'}_{i}^{1}...{b'}_{i}^{Z-1}{n'}_{i}{b}_{i}^{1}...{b}_{i}^{Z-1}{n}_{i}} \\
&=[\hat{O}_{\alpha}]^{{b'}_{i}^{\alpha}{b }_{i}^{\alpha}}
  [\hat{O}_{\beta }]^{{b'}_{i}^{\beta }{b }_{i}^{\beta }}
  [\hat{O}_{\gamma}]^{{b'}_{i}^{\gamma}{b }_{i}^{\gamma}} \\
& \quad \times \hat{P}(\bra{{b'}_{i}^{\alpha+1}}...\bra{{b'}_{i}^{Z-1}}\bra{{n'}_{i}},\hat{O}_{\alpha}) \\
& \quad \times \hat{P}(\bra{{b'}_{i}^{\beta +1}}...\bra{{b'}_{i}^{Z-1}}\bra{{n'}_{i}},\hat{O}_{\beta }) \\
& \quad \times \hat{P}(\bra{{b'}_{i}^{\gamma+1}}...\bra{{b'}_{i}^{Z-1}}\bra{{n'}_{i}},\hat{O}_{\gamma}),
\end{split}
\end{equation}

\begin{equation}
\label{eq:Ops4}
\begin{split}
& [\hat{O}_{\alpha} \hat{O}_{\beta } \hat{O}_{\gamma} \hat{O}_{\delta}]^{{b'}_{i}^{1}...{b'}_{i}^{Z-1}{n'}_{i}{b}_{i}^{1}...{b}_{i}^{Z-1}{n}_{i}} \\
&=[\hat{O}_{\alpha}]^{{b'}_{i}^{\alpha}{b }_{i}^{\alpha}}
  [\hat{O}_{\beta }]^{{b'}_{i}^{\beta }{b }_{i}^{\beta }}
  [\hat{O}_{\gamma}]^{{b'}_{i}^{\gamma}{b }_{i}^{\gamma}}
  [\hat{O}_{\delta}]^{{b'}_{i}^{\delta}{b }_{i}^{\delta}} \\
& \quad \times \hat{P}(\bra{{b'}_{i}^{\alpha+1}}...\bra{{b'}_{i}^{Z-1}}\bra{{n'}_{i}},\hat{O}_{\alpha}) \\
& \quad \times \hat{P}(\bra{{b'}_{i}^{\beta +1}}...\bra{{b'}_{i}^{Z-1}}\bra{{n'}_{i}},\hat{O}_{\beta }) \\
& \quad \times \hat{P}(\bra{{b'}_{i}^{\gamma+1}}...\bra{{b'}_{i}^{Z-1}}\bra{{n'}_{i}},\hat{O}_{\gamma}) \\
& \quad \times \hat{P}(\bra{{b'}_{i}^{\delta+1}}...\bra{{b'}_{i}^{Z-1}}\bra{{n'}_{i}},\hat{O}_{\delta}).
\end{split}
\end{equation}

Here, the representation of operator $[\hat{O}]^{{b'}_{i}^{1}...{b'}_{i}^{Z-1}{n'}_{i}{b}_{i}^{1}...{b}_{i}^{Z-1}{n}_{i}}$ is
a $4M^{Z} \times 4M^{Z}$ matrix.
In the renormalization step, this is reduced to $M \times M$ matrix $[\hat{O}]^{{b'}_{i}^{Z}{b}_{i}^{Z}}$,
spanned by renormalized states $\ket{b_{i}^{Z}}$ using a tensor $\mathbf{A}^{n_{i}}$ as,

\begin{equation}
\label{eq:decimate}
   \sum_{\substack{{b'}_{i}^{1}...{b'}_{i}^{Z-1}{n'}_{i}\\{b}_{i}^{1}...{b}_{i}^{Z-1}{n}_{i}}}
  A_{{b'}_{i}^{1}...{b'}_{i}^{Z}}^{{n'}_{i}\dag}
  [\hat{O}]^{{b'}_{i}^{1}...{b'}_{i}^{Z-1}{n'}_{i}{b}_{i}^{1}...{b}_{i}^{Z-1}{n}_{i}}
  A_{{b }_{i}^{1}...{b }_{i}^{Z}}^{{n }_{i}}.
\end{equation}
It is necessary to minimize the cost of tensor contractions for TTNS renormalization (\ref{eq:decimate}) is just as for $\hat{H}\psi$, as described in the text.
To compute representations of complementary operators, each complementary operator can be further divided into $Z-1$ renormalized blocks and the site \textit{i} itself.
We carefully minimized these tensor contraction costs for each complementary operator and their complexities per site are summarized in Table~\ref{tbl:renormalization} for a general tree.
Thus, for a general tree, the computational complexity of a TTNS sweep is $\mathcal{O}(M^{Z+1}k^{3} + M^{Z}k^{5})$.
As discussed in the text, the complexity can be further reduced to $\mathcal{O}(M^{3}k^{3}+M^{2}k^{4})$ for $Z=2$ (MPS) and
$\mathcal{O}(M^{4}k^{3}+M^{2}k^{5})$ for $Z=3$.

\begin{table}
\begin{center}
\caption{Complexity of optimal tensor contractions for TTNS renormalization per site. Multiplying by $\mathcal{O}(k)$ gives complexity per sweep.}
\label{tbl:renormalization}
\begin{ruledtabular}
\begin{tabular}{ c l }
  complementary operator            & complexity \\
\hline
  $\hat{H}$                         & $\mathcal{O}(M^{Z+1}k^{2}+M^{Z}k^{4})$ \\
  $\hat{a}_{i}$                     & $\mathcal{O}(M^{Z+1}k)$ \\
  $\hat{S}_{i}$                     & $\mathcal{O}(M^{Z+1}k^{2}+M^{Z}k^{4})$ \\
  $\hat{a}_{i}\hat{a}_{j}$          & $\mathcal{O}(M^{Z+1}k^{2})$ \\
  $\hat{a}_{i}^{\dag}\hat{a}_{j}$   & $\mathcal{O}(M^{Z+1}k^{2})$ \\
  $\hat{P}_{ij}$                    & $\mathcal{O}(M^{Z+1}k^{2}+M^{2}k^{4})$ \\
  $\hat{Q}_{ij}$                    & $\mathcal{O}(M^{Z+1}k^{2}+M^{2}k^{4})$ \\
\end{tabular}
\end{ruledtabular}
\end{center}
\end{table}


\end{document}